\documentclass[journal]{IEEEtran}
\usepackage[linesnumbered,noend]{algorithm2e}

\usepackage{psfrag}

\usepackage{cite}

\ifCLASSINFOpdf
\else
  \usepackage[dvips]{graphicx}
  \graphicspath{{figures/}}
  \DeclareGraphicsExtensions{.eps}
\fi
\usepackage[cmex10]{amsmath}
\usepackage[caption=false,font=footnotesize]{subfig}

\newtheorem{theorem}{Theorem}

\begin{document}
\newdimen\snellbaselineskip
\newdimen\snellskip
\snellskip=1.5ex
\snellbaselineskip=\baselineskip
\def\srule{\omit\kern.5em\vrule\kern-.5em}
\newbox\bigstrutbox
\setbox\bigstrutbox=\hbox{\vrule height14.5pt depth9.5pt width0pt}
\def\bigstrut{\relax\ifmmode\copy\bigstrutbox\else\unhcopy\bigstrutbox\fi}
\def\middlehrule#1#2{\noalign{\kern-\snellbaselineskip\kern\snellskip}
&\multispan#1\strut\hrulefill
&\omit\hbox to.5em{\hrulefill}\vrule 
height \snellskip\kern-.5em&\multispan#2\hrulefill\cr}

\makeatletter
\def\bordermatrix#1{\begingroup \m@th
  \@tempdima 8.75\p@
  \setbox\z@\vbox{%
    \def\cr{\crcr\noalign{\kern2\p@\global\let\cr\endline}}%
    \ialign{$##$\hfil\kern2\p@\kern\@tempdima&\thinspace\hfil$##$\hfil
      &&\quad\hfil$##$\hfil\crcr
      \omit\strut\hfil\crcr\noalign{\kern-\snellbaselineskip}%
      #1\crcr\omit\strut\cr}}%
  \setbox\tw@\vbox{\unvcopy\z@\global\setbox\@ne\lastbox}%
  \setbox\tw@\hbox{\unhbox\@ne\unskip\global\setbox\@ne\lastbox}%
  \setbox\tw@\hbox{$\kern\wd\@ne\kern-\@tempdima\left(\kern-\wd\@ne
    \global\setbox\@ne\vbox{\box\@ne\kern2\p@}%
    \vcenter{\kern-\ht\@ne\unvbox\z@\kern-\snellbaselineskip}\,\right)$}%
  \null\;\vbox{\kern\ht\@ne\box\tw@}\endgroup}
\makeatletter

\makeatletter
\def\bordermatrix#1{\begingroup \m@th
  \@tempdima 8.75\p@
  \setbox\z@\vbox{%
    \def\cr{\crcr\noalign{\kern2\p@\global\let\cr\endline}}%
    \ialign{$##$\hfil\kern2\p@\kern\@tempdima&\thinspace\hfil$##$\hfil
      &&\quad\hfil$##$\hfil\crcr
      \omit\strut\hfil\crcr\noalign{\kern-\snellbaselineskip}%
      #1\crcr\omit\strut\cr}}%
  \setbox\tw@\vbox{\unvcopy\z@\global\setbox\@ne\lastbox}%
  \setbox\tw@\hbox{\unhbox\@ne\unskip\global\setbox\@ne\lastbox}%
  \setbox\tw@\hbox{$\kern\wd\@ne\kern-\@tempdima\left(\kern-\wd\@ne
    \global\setbox\@ne\vbox{\box\@ne\kern2\p@}%
    \vcenter{\kern-\ht\@ne\unvbox\z@\kern-\snellbaselineskip}\,\right)$}%
  \null\;\vbox{\kern\ht\@ne\box\tw@}\endgroup}
\makeatletter

%
\title{Towards a Collision-Free WLAN: Dynamic Parameter Adjustment in CSMA/E2CA}
%
%
%

\author{
Jaume Barcelo, Boris Bellalta, Cristina Cano, Anna Sfairopoulou, Miquel Oliver and Kshitiz Verma
\thanks{J. Barcelo is with Universidad Carlos III de Madrid.}%
\thanks{B. Bellalta, C. Cano, A. Sfairopoulou and M. Oliver are with Universitat Pompeu Fabra.}%
\thanks{K. Verma is with Universidad Carlos III de Madrid and Institute IMDEA Networks.}%
\thanks{Correspondence should be addressed to Boris Bellalta, Carrer Tanger 122-140, 08018 Barcelona. E-mail: boris.bellalta@upf.edu}
}

%
%

\markboth{}%
{Towards a Collision-Free WLAN: Dynamic Parameter Adjustment in CSMA/E2CA}
%



\maketitle

\begin{abstract}
Carrier Sense Multiple Access with Enhanced Collision Avoidance (CSMA/ECA) is a distributed MAC protocol that allows collision-free access to the medium in WLAN. The only difference between CSMA/ECA and the well-known CSMA/CA is that the former uses a deterministic backoff after successful transmissions. Collision-free operation is reached after a transient state during which some collisions may occur. This article shows that the duration of the transient state can be shortened by appropriately setting the contention parameters. Standard absorbing Markov Chain theory is used to describe the behaviour of the system in the transient state and to predict the expected number of slots to reach the collision-free operation. 

The article also introduces CSMA/E2CA, in which a deterministic backoff is used two consecutive times after a successful transmission. CSMA/E2CA converges quicker to collision-free operation and delivers higher performance than CSMA/ECA, specially in harsh wireless scenarios with high frame error rates.

The last part of the article addresses scenarios with a large number of contenders. We suggest dynamic parameter adjustment techniques to accommodate a varying (and potentially high) number of contenders. The effectiveness of these adjustments in preventing collisions is validated by means of simulation.


\end{abstract}


%
\IEEEpeerreviewmaketitle

\section{Introduction}
%
%
%
%

\IEEEPARstart{W}{ireless} Local Area Networks (WLAN) are a popular choice as a last-hop connection to the Internet. Many consumer devices support the IEEE 802.11 standard \cite{IEEE80211-IEEESTD2007} and share the radio channel to access the Internet wirelessly. As the number of wireless devices with data to transmit increases, the likelihood that two or more of these devices simultaneously transmit also increases. If a collision between two simultaneous transmissions occurs, the intended receiver of the transmission may not be able to decode the information that has been sent. In some practical scenarios such as conference rooms, the number of collisions may render the wireless network unusable. This paper explores an approach that can reduce the number of collisions and may even allow for collision-free operation of the network.

In the remainder of the article, we assume the familiarity of the reader with the IEEE 802.11 standard and the related terminology. The focus of the article is on the medium access control (MAC) layer of the protocol and, in particular, on the distributed coordination function (DCF) and the more recent enhanced distributed channel access (EDCA). Both approaches heavily rely on carrier sense multiple access with collision avoidance (CSMA/CA). This has proven to be a lightweight, robust, and effective protocol in infrastructure scenarios in which a number of nearby stations connect to an access point for Internet access.

A downside of CSMA/CA is that only a fraction of the channel time is devoted to successful transmissions while the remainder is wasted in the form of collisions and empty channel. Sec. \ref{sec:related_work} reviews previous research efforts aimed to increase the efficiency of the protocol, i.e., the fraction of channel time devoted to successful transmissions. The emphasis is placed on CSMA with Enhanced Collision Avoidance (CSMA/ECA), a subtle modification to CSMA/CA that uses a deterministic backoff after successful transmissions to avoid collisions. CSMA/ECA reaches collision-free operation after a transient state. We proceed in Sec. \ref{sec:analysis} by showing that, if the minimum and maximum contention window are set to the same value, the duration of the transient state is reduced. The transition to collision-free operation can be modelled as an absorbing Markov Chain. Standard Markov Chain theorems can be used to estimate the expected number of slots required to reach the collision-free operation. These analytical results are validated by means of simulation. Sec. \ref{sec:e2ca} suggests using a deterministic backoff for two consecutive times after a successful transmission. This variation is called CSMA/E2CA and we study the absorption process to show that its quicker than in the original CSMA/ECA. More importantly, CSMA/E2CA proves to be more resilient in harsh wireless conditions where the frame error rates are high. A limitation of CSMA/E2CA is that the number of contenders that can operate in a collision-free fashion is limited. Sec. \ref{sec:dpa} deals with dynamic parameter adjustment in a CSMA/E2CA network to accommodate a varying (and potentially large) number of contenders. The possibility of using a deterministic backoff for several consecutive times after a successful transmission in briefly discussed in Sec. \ref{sec:stickiness}. Finally, Sec. \ref{sec:conclusion} concludes the article.


\section{Related Work}
\label{sec:related_work}
Since Abramson's team seminal design and deployment of the Aloha wireless network \cite{abramson1970asa,abramson2009asw}, the MAC protocols have evolved to achieve higher efficiency, which is defined as the fraction of channel time devoted to successful transmissions. By dividing channel time into slots, Slotted Aloha may double the efficiency of the original protocol under certain conditions (such as fixed packet length equal to the slot size) \cite{rom1990map,roberts1975aps,tanenbaum1988cn}. In Reservation-Aloha \cite{crowther1973sbc}, a station that successfully transmits obtains a reservation for a slot that follows a deterministic number of slots later. 

In WLANs, it is possible to substantially reduce the duration of empty slots thanks to the fact that the propagation times are short and thus we can quickly determine that a slot is empty by simply sensing the channel. Shortening the duration of empty slots increases the efficiency of the MAC protocol. This is the idea underlying CSMA/CA. An additional performance improvement can be obtained when the aggressiveness of the contending stations is adjusted as a function of the number of contending stations \cite{cali2000dti}. Channel observation and advanced filtering techniques can be used to obtain an accurate estimate of the number of contenders and adjust the contention parameters accordingly \cite{toledo2006aoi}. It is even possible to get close to the optimal efficiency of CSMA/CA without any knowledge of the number of contenders. Since optimal collision probability is an invariant that does not depend on the number of contenders, control theory and a feedback loop can be used to adjust the contention parameters \cite{patras2009cta}.

Unfortunately, the optimal efficiency of CSMA/CA is still far from one. In other words, a substantial amount of channel time is still wasted in the form of empty slots and collisions. One of the latest research trends in MAC protocols for WLANs is the use of learning protocols. These protocols are executed in a distributed fashion but they are not completely random. On the contrary, they try to gather information about other stations' intentions to transmit in order to schedule their own transmissions in a way that reduces the chances of collisions. A reduction in the number of collisions results in significant performance gains, and these protocols may even achieve collision-free operation.

A first example of these protocols is the Enhanced Backoff Algorithm (EBA) introduced in \cite{choi2005eei}, which proposes that each station announces its backoff intentions in a special header to avoid collisions. A caveat is that currently deployed hardware would discard EBA packets since these are not IEEE 802.11 compliant and thus backward compatibility is compromised.

If the stations use a deterministic backoff after successes, it is not longer required to include a header explicitly stating the backoff intentions. Moreover, when all the stations use the same deterministic backoff after successes, the system naturally converges to a collision-free operation in which the stations transmit in a round-robin deterministic fashion. We call this variant of CSMA/CA that uses a deterministic backoff after successes CSMA/ECA\footnote{This idea was initially presented with the name of Learning Binary Exponential Backoff or L-BEB. Subsequent discussion with the community convinced the authors that the idea was more related to avoiding collisions than to doubling contention windows. Thus the protocol was re-named Carrier Sense Multiple Access with Enhanced collision Avoidance or CSMA/ECA. The authors apologise for the confusion derived from the change of the name.} \cite{barcelo2008lba}.

ZeroCollision \cite{lee2008daa} is a MAC protocol that also converges to a collision-free mode of operation. The operation of ZeroCollision is similar to Reservation-Aloha. The slots are grouped in rounds, each round containing $C$ slots and the stations keep track of the status of each slot (either busy or empty). A station that successfully transmits in the $c$-th slot ($0<c<C$) obtains an implicit reservation for that slot. If this station has more packets to transmit, it will use again the $c$-th slot in the next transmission round. Those stations that have no reservation, transmit randomly in any of the slots that were empty in the previous round.

An insightful comparison between CSMA/ECA and ZeroCollision is presented in \cite{fang2009dlm}. ZeroCollision converges quicker to the collision-free operation and CSMA/ECA presents a behaviour that is more similar to current CSMA/CA. This similarity may ease the implementation of CSMA/ECA. The results in this reference also show that ZeroCollision is superior to CSMA/ECA in the presence of channel errors. Additionally, the cited article also extends both protocols to converge quicker to collision-free operation, to deliver higher performance in lossy channels, and to operate in a collision-free fashion in the presence of a large number of stations. The names of these extended protocols are L-ZC, L-MAC, A-L-ZC and A-L-MAC. Furthermore, the concept of \emph{stickiness} is introduced. We will revisit this concept in Section \ref{sec:e2ca} as a means to substantially improve the performance of CSMA/ECA.

\begin{figure*}%
  \centering
  \includegraphics[width=2.0in]{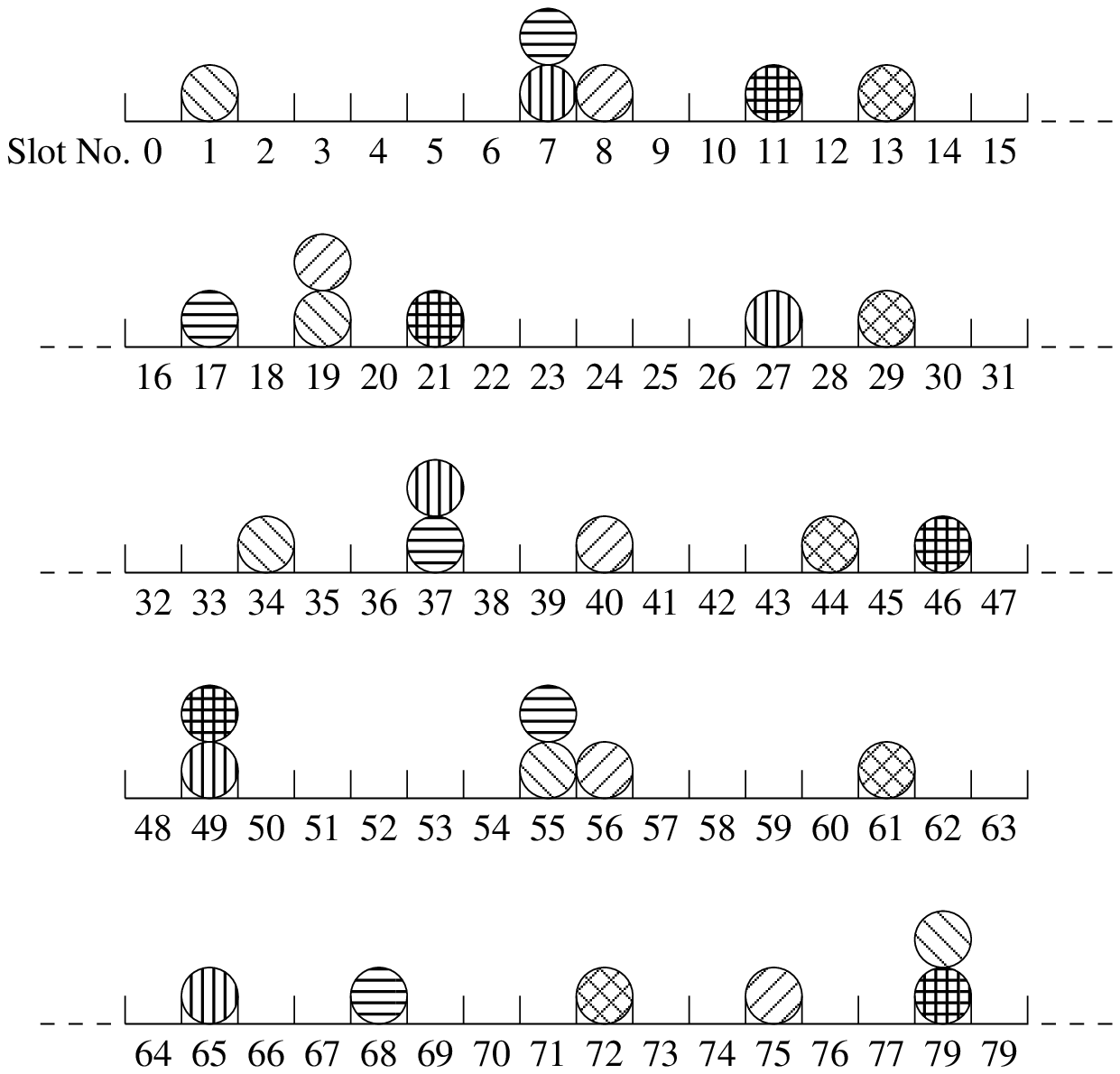}
  \qquad
  \includegraphics[width=2.0in]{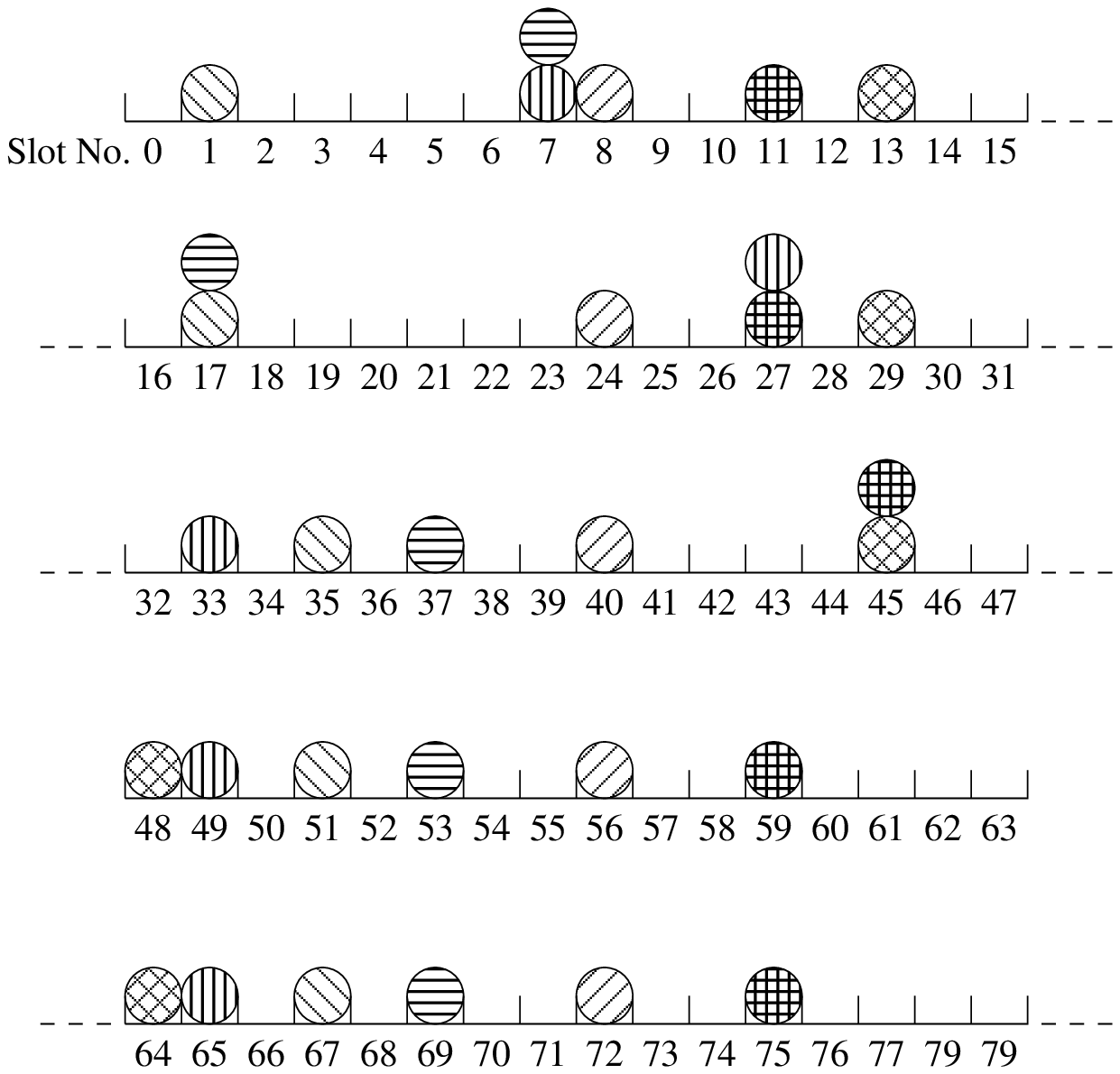}
  \qquad
  \includegraphics[width=2.0in]{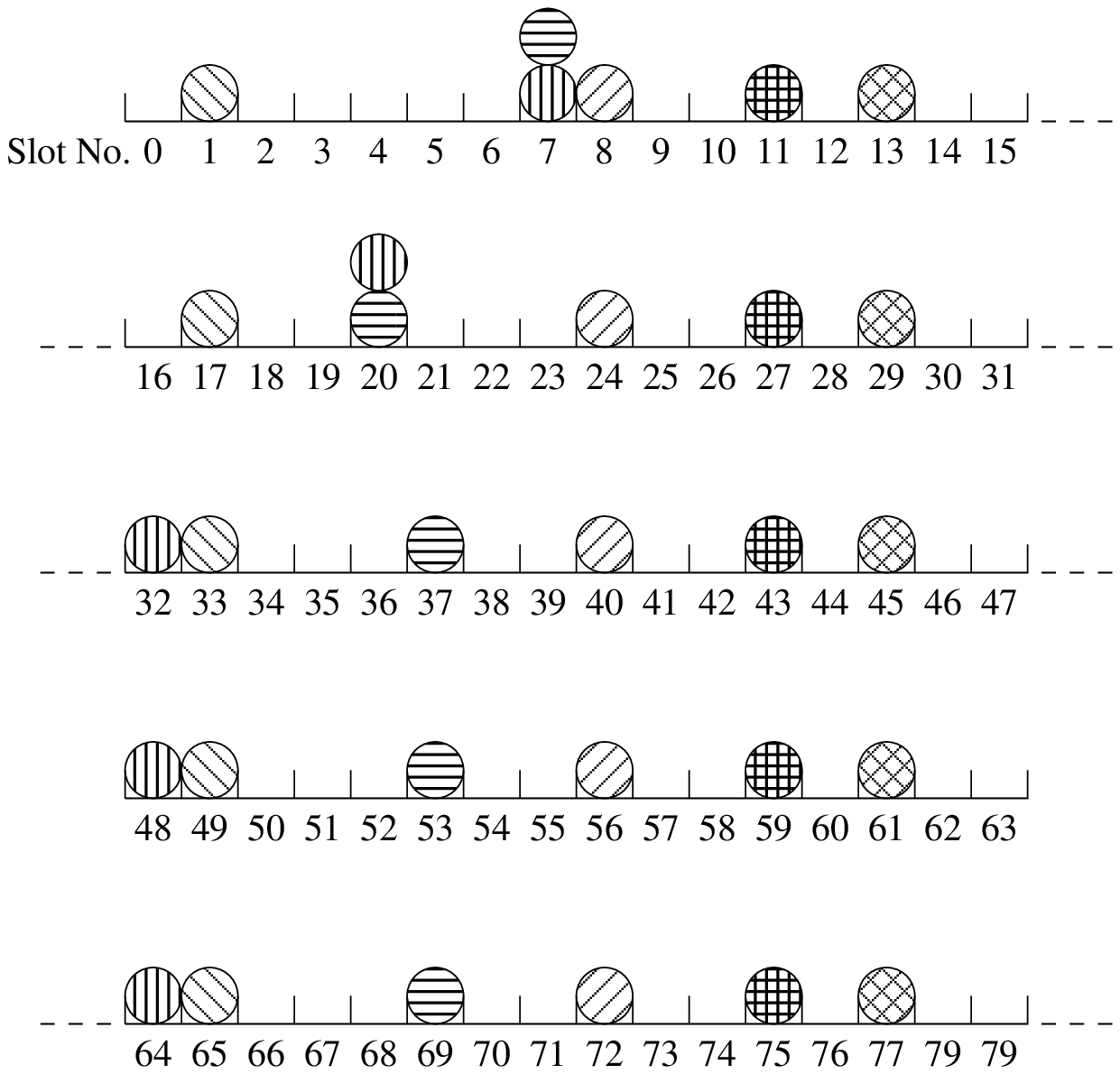}
  \caption{An example of CSMA/CA (left), CSMA/ECA (center) and ZeroCollision (right) execution.}%
  \label{fig:contention_example}%
\end{figure*}

Fig. \ref{fig:contention_example} provides an example of the operation of CSMA/CA, CSMA/ECA and ZeroCollision. In each case there are six different stations contending for the channel. A transmission by one of the stations is represented as a ball in the figure. We use different filling patterns for balls belonging to different stations. The bins represent the slots. Even though all bins are depicted as being equal for convenience, busy slots are substantially longer than empty slots in reality.

In the figure, it can be observed that CSMA/CA uses a random backoff before each transmission attempt and it never converges to collision-free operation. CSMA/ECA uses a deterministic backoff after successful transmissions and a random backoff otherwise. The value of the deterministic backoff is constant for all stations, $C=16$, which is equal to the expectation of the random backoff which is used in CSMA/CA. Note that $C$ is also the maximum number of stations that can contend in a collision-free fashion. For this reason we sometimes refer to $C$ as the capacity of the system.

In the figure, we can see that at some point all the CSMA/ECA stations successfully transmit and the system operates in a deterministic and collision-free fashion. A property of CSMA/ECA is that a station that has successfully transmitted in its last transmission attempt never collides with another station that has also succeeded in its last transmission attempt.

In our ZeroCollision example, we also use a schedule size $C=16$ in order to ease comparison with CSMA/CA. In ZeroCollision, a station that successfully transmits does not collide in its next transmission attempt. As a result, ZeroCollision converges quicker to collision-free operation.

The idea of using a deterministic backoff after successful transmissions is also presented in \cite{he2009sbr}, where it is called Semi-Random Backoff (SBR). This reference studies the convergence to collision-free operation, the collision probability and the performance in presence of hidden terminals. Delay measures are also presented and many other performance issues are explored. It is concluded that SBR performs equal or better than the random backoff in all possible scenarios.

The focus of the present paper is on single-hop WLAN communications. The broad idea of learning MAC protocols, stickiness and convergence to a collision-free schedule has been studied in the context of wireless multi-hop networks in \cite{yi2010msl,singh2007scc,lin1997amm}.

The remainder of this paper deals with CSMA/ECA. We have studied several aspects of CSMA/ECA in our previous work. In particular, the performance of CSMA/ECA for rigid flows, elastic flows, and a combination of both is assessed by means of simulation in \cite{bellalta2009vtc}. An analytical model that captures the behaviour of the protocol for both saturated and non-saturated scenarios is presented in \cite{barcelo2009cpa}. The efficacy of CSMA/ECA in providing traffic differentiation is demonstrated in \cite{barcelo2009tpc}. In \cite{barcelo2010fcc}, the coexistence of CSMA/ECA with the legacy protocol is studied. This last reference also presents a preliminary study of the duration of the transient state. Finally, a possible solution to the problem of accommodating a large number of stations in a collision-free fashion in infrastructure WLAN is presented in \cite{barcelo2010dpa}. The present article, reviews and substantially extends \cite{barcelo2010dpa}.

In this article, unless otherwise stated, we will rely on the following assumptions:
\begin{itemize}
\item Ideal channel that does not introduce errors.
\item There is no capture effect, i.e., a collision always results in failure.
\item All the stations are in the same collision domain and we have perfect carrier sensing.
\item All the stations have the same priority.
\item All the stations are saturated, which means that they always have a packet ready to be transmitted.
\item All the stations are synchronized to the channel slots. We do not take into account clock drifts.
\end{itemize}

The current article presents the following contributions that extend and complement previous work:
\begin{itemize} 
\item We study the particular case in which the minimum contention window is equal to the maximum contention window ($CW_{min}=CW_{max}$). We show that this configuration reduces the duration of the transient state.
\item We make use of standard Absorbing Markov Chain theory to estimate the average number of slots required to reach collision-free operation.
\item We suggest the use of a deterministic backoff for two consecutive times after a successful transmission. In other words, the stations will keep using a deterministic backoff until two consecutive packet losses occur. We call the protocol CSMA/E2CA and we show that this approach further reduces the duration of the transient state.
\item We show that CSMA/E2CA clearly outperforms CSMA/CA and CSMA/ECA in scenarios in which packet losses occur.
\item We revisit the idea of dynamic parameter adjustment initially presented in \cite{barcelo2010dpa}. We enhance it by incorporating an hysteresis margin to prevent the oscilation of the contention parameter and we propose an alternative implementation using integer arithmetic. 
\end{itemize}

\section{Parameter Setting and Analytical Model}
\label{sec:analysis}
CSMA/ECA is an attractive contention protocol because it is very similar to the pervasive CSMA/CA and can achieve collision-free operation. In \cite{barcelo2010fcc} it is explained that the pseudocode description of CSMA/ECA and CSMA/CA differ in a single line. This makes it very easy to try CSMA/ECA in any simulator and in principle it should be also easy for the manufacturers to switch to the new protocol. The fact that the two protocols are so similar, allows for the smooth coexistence of CSMA/ECA and CSMA/CA station and would allow CSMA/ECA devices to pass WiFi certification tests.

The value of the deterministic backoff after successes in CSMA/ECA is computed as a function of the minimum contention window ($CW_{min}$) as
\begin{equation}
C = \lceil E \left[ \mathcal{U} [0,CW_{min} - 1 ] \right] \rceil,
\label{eq:deterministic_backoff}
\end{equation}

where $\lceil \cdot \rceil$ is the ceiling operator, $E \left[ \cdot \right]$ is the expectation operator and $\mathcal{U}$ is the uniform distribution. In our previous work we have already shown that using the parameter adjustment possibilities of the latest version of the standard, it is possible to attain traffic differentiation \cite{barcelo2009tpc} and accommodate a large number of contenders \cite{barcelo2010dpa} by adjusting $CW_{min}$.
In the following we will show that we can reduce the time required to reach collision free operation by using a smaller maximum contention window, in particular $CW_{max}=CW_{min}$. 
Since the transient state can be seen as an stochastic (trial-and-error) search for a collision free schedule, setting $CW_{max} = CW_{min}$ can be intuitively interpreted as reducing the time between trials which results in a faster search. We will show that this is indeed the case.

The doubling of the contention window after each unsuccessful transmission is called Binary Exponential Backoff (BEB) and its purpose is to reduce the transmission probability of the stations when collisions occur. When the number of stations is large, the role of BEB is to reduce the aggressiveness of the stations to reduce the number of collisions. It was pointed out in \cite{banchs2006tao} that when there are alternative means available to adjust the contention window, BEB simply decreases the fairness and the overall efficiency of a IEEE 802.11e network. The same argument applies if CSMA/ECA is used.

In a CSMA/CA+BEB network (e.g., a IEEE 802.11 network), a station that collides is doubly penalized. First its payload is not delivered, and second it has to wait for a longer time before the next transmission attempt. In CSMA/ECA+BEB, a station that collides is triply penalized, because the stations that randomly select their backoff perceive a higher collision probability than those that select their backoff deterministically.

\begin{figure}[!t]
\centering
\includegraphics[width=3.5in]{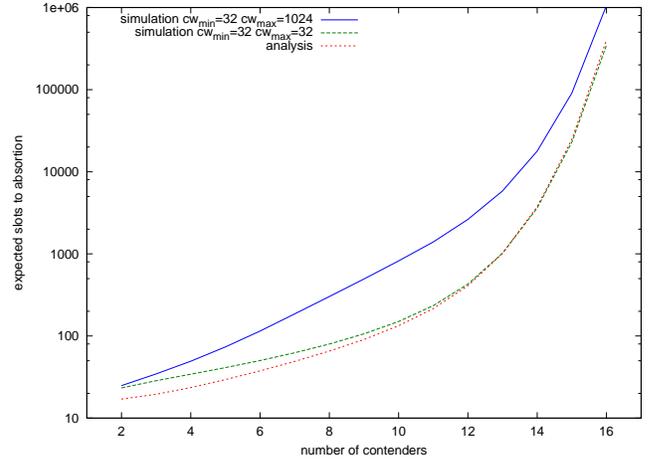}
\caption{Number of expected slots to reach collision-free operation in CSMA/ECA.}
\label{fig:absorption}
\end{figure}

Furthermore, BEB unnecessarily increases the number of slots required for the system to reach collision-free operation. This can be readily observed in Fig. \ref{fig:absorption} that depicts the number of slots required for the system to converge when a given number of stations simultaneously join the contention. To obtain this results we have used a minimum contention window $CW_{min}=32$ and a deterministic backoff after successes $C=16$. The number of contending stations $\sigma$ ($2\leq \sigma \leq C$) is represented in the $x$ axis. 

The first two curves are obtained by means of simulation\footnote{We have used custom simulators in C. The simulators only implement the MAC layer and assume that the stations are saturated and in the same collision domain. The channel is ideal an does not introduce errors. The presented results are averages of $10^4$ runs. Source code is available upon request to the first author.}. In the first one we use different values for the minimum and maximum contention window ($CW_{min}=32$ and $CW_{max}={1024}$) while in the second one the same value is used for both contention windows ($CW_{min} = CW_{max} = 32$). It can be observed that using the same value reduces the number of slots required for the system to reach collision-free operation. In the remainder of the article, unless otherwise stated, we will not use BEB (e.g., we will set the maximum contention window to the same value of the minimum contention window).

When the minimum and maximum contention windows are equal, the system can be modeled as an absorbing Markov Chain and it is possible to predict analytically the number of slots required to reach the collision-free operation.

To model the system we divide the sequence of slots in groups of $C$ slots and we will refer to each group as a step. This is exactly what we have done in Fig. \ref{fig:contention_example} to ease representation. In the figure, slots ranging from 0 to 15 belong to the first step, and those between 16 and 31 belong to the second step. We assume that all the stations transmit once in each step (this is a modelling approximation, since it is not satisfied in the real system). Furthermore, we define the state of the system as the number of successful transmissions in each step.

 Then we can model the transition process as an absorbing time-homogeneous Markov Chain whose state space is 
\begin{equation}
\mathbf{S} =\{S_i | 0 \leq i \leq \sigma \},
\end{equation}
with initial state $S_0$ and a stable state $S_\sigma$, where $\sigma$ is the number of contenders.

The details to compute the transition matrix $\mathbf{P}$ of this Markov Chain are provided in \cite{barcelo2010fcc}. We reproduce here a simple example from that paper in which the number of contenders is $\sigma=3$ and the deterministic backoff after successes is $C=4$.

\begin{equation}
\mathbf{P}_{\sigma=3,C=4} =  
\left( \begin{array}{cccc}
 \frac{1}{16} & \frac{9}{16} & 0 & \frac{6}{16} \\
 \frac{1}{16} & \frac{9}{16} & 0 & \frac{6}{16} \\
 0            & \frac{1}{2}  & 0 & \frac{1}{2} \\
 0            & 0            & 0 & 1 
\end{array} \right)
\end{equation}

$\mathbf{P}$ is a square matrix of size $\sigma + 1$. The last row contains only zeros with the exception of the last column which is one. This property characterizes the system as an absorbing Markov Chain. It means that when the last state $S_\sigma$ is reached, the system transits to $S_\sigma$ with probability 1, i.e., the state $S_\sigma$ is absorbing.

In the following, we will closely follow the discussion on absorbing Markov Chains presented in \cite{grinstead1997itp}. The proofs of the theorems we need can be found in Chapter 11 of this reference.

The canonical representation of our transition matrix is the following.

\begin{equation}
\offinterlineskip
\mathbf{P}\;= \bordermatrix{      
                               &\hbox{TR.}&\omit\hfil&\hbox{ABS.}\cr
           \hbox{TR.}\bigstrut &\mathbf{Q}   &\srule    &\mathbf{R}    \cr
\middlehrule{1}{1}
           \hbox{ABS.}\bigstrut&\mathbf{0}   &\srule    &\mathbf{1}}
\end{equation} 

where $\mathbf{1}$ is a $1$-by-$1$ identity matrix, $\mathbf{0}$ is a $1$-by-$\sigma$ zero matrix, $\mathbf{R}$ is a nonzero $\sigma$-by-$1$ matrix, and $\mathbf{Q}$ is a $\sigma$-by-$\sigma$ matrix. The rows and columns that belong to the transient (TR.) and absorbing (ABS.) states are indicated in the equation.

The following theorem guarantees that the collision-free mode of operation will be eventually reached.

\begin{theorem}
In an absorbing Markov chain, the probability that the process will be absorbed is 1.

\end{theorem}

Then we compute the fundamental matrix $\mathbf{N}$ for our system.
\begin{theorem}
For an absorbing Markov chain the matrix \mbox{$\mathbf{I} - \mathbf{Q}$} has an inverse
$\mathbf{N}$ and 
$\mathbf{N}  =\mathbf{I} + \mathbf{Q} + \mathbf{Q}^{2} + \cdots\ $.  The $ij$-entry
$n_{ij}$ of the 
matrix $\mathbf{N}$ is the expected number of times the chain is in state $s_j$,
given that 
it starts in state $s_i$.  The initial state is counted if $i = j$.
\end{theorem}

The relationship between the fundamental matrix $\mathbf{N}$ and the number of steps to absorption is stated in the next theorem. 

\begin{theorem} Let $t_i$ be the expected number of steps
before
the chain is absorbed, given that the chain starts in state $s_i$, and let
$\mathbf{t}$ 
be the column vector whose $i$th entry is $t_i$. Then
$$\mathbf{t} = \mathbf{N}\mathbf{c}\ ,$$
where $\mathbf{c}$ is a column vector all of whose entries are 1.
\end{theorem}

Now $t_1$ is the expected number of steps to absorption assuming that all the stations simultaneously join the contention. To obtain the expected number of slots to absorption we have to multiply the number of steps by $C$, which is the number of slots in each step. In Fig. \ref{fig:absorption}, these analytical results are compared with the simulation results. It can be observed that our model can be used to predict the expected duration of transient state when $CW_{min}=CW_{max}$. 

There is a small gap between the analytical and simulation curve in Fig.\ref{fig:absorption} when the number of slots to absorption is low. The explanation lies in the aforementioned modeling assumption that all the stations transmit exactly once in each of the steps. The results that we obtained are in agreement with the ones presented in \cite{he2009sbr} in Table II.

\section{CSMA/E2CA and performance in the presence of errors}
\label{sec:e2ca}

One of the shortcomings of CSMA/ECA when compared to ZeroCollision is that it takes longer for CSMA/CA to reach the collision-free operation. In this section we suggest a simple idea to dramatically accelerate the convergence of CSMA/ECA. 

In ZeroCollision, a station that successfully transmits in its last transmission attempt will not collide in its next transmission attempt. The reason is that every station keeps track of the other station's successful transmissions and intelligently schedules the next transmission attempt to avoid a collision with those stations that deterministically choose their transmission slot for their next transmission attempt.

Our aim is to shorten the transient state without requiring that each station keeps track of the other stations' successes. Our idea relies on the concept of stickiness introduced in \cite{fang2009dlm}. This reference suggests that a station \emph{sticks} (with a given probability) to a given slot after suffering a collision. Our proposal is that a station chooses a deterministic backoff for two consecutive times after a successful transmission. Then a single collision (or channel error) is not enough to move a station from the deterministic mode of operation to the random mode of operation. Suffering two consecutive collisions or channel errors is an unlikely event.

\begin{figure}[!t]
\centering
\includegraphics[width=2.0in]{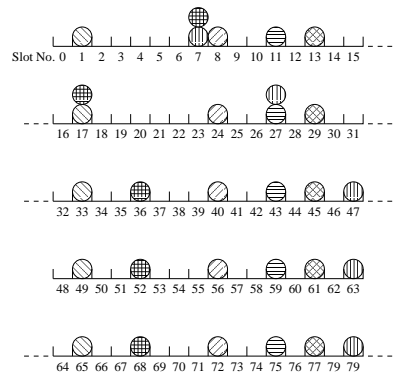}
\caption{A graphical example of the execution of CSMA/E2CA}
\label{fig:csma_e2ca}
\end{figure}

We call the protocol that uses a deterministic backoff for two consecutive times after each successful transmission CSMA/E2CA. This behaviour is exemplified in Fig. \ref{fig:csma_e2ca}. Take as an example the transmission in slot 11 which is a ball filled with horizontal lines. This station successfully transmits and then it suffers a collision in its next transmission attempt in slot 27. Despite this collision, the station chooses a deterministic backoff again. 
 In our example, the station succeeds in the next transmission attempt in slot 43 and therefore it keeps using a deterministic backoff.
Had the station suffered a new collision in the next transmission attempt in slot 43, it would have used a random backoff in the following transmission attempt. In CSMA/E2CA, a station that suffers two consecutive unsuccessful transmission attempts switches to the random mode of operation.

\begin{figure}[!t]
\centering
\includegraphics[width=3.5in]{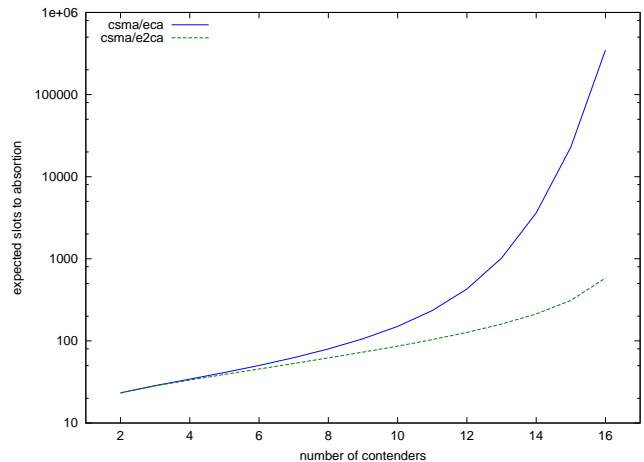}
\caption{Expected number of slots required to reach collision-free operation in CSMA/ECA and CSMA/E2CA for various numbers of contenders.}
\label{fig:absorption_e2ca}
\end{figure}

The stickiness of CSMA/E2CA allows for a quicker convergence to collision-free operation as can be observed in Fig. \ref{fig:absorption_e2ca}. This figure compares simulation results for the average number of slots to absorption in CSMA/ECA and CSMA/E2CA. Stickiness can reduce the expected number of slots to convergence by orders of magnitude when the number of contenders $\sigma$ has a value that is close or equal to the capacity $C$.

In the remainder of this section we will compare four protocols: namely CSMA/CA, CSMA/CA + BEB, CSMA/ECA and CSMA/E2CA. In a first round of simulation we will use an ideal channel that does not drop packets. This will set the upper bound for the performance of each of the protocols. We opted for measuring the number of empty, success and collision slots. These measures can be translated to throughput for a given IEEE 802.11 physical flavor (e.g., 11a, 11b, 11g or 11n) and a given packet length distribution.

The simulations last for $10^6$ slots and the transitory is taken into account. This is an important detail since performance in the transient state is traded off for performance in the steady state. Those simulations with a large number of contenders present a higher performance in the steady-state but are penalized by longer transient states.

\begin{figure}[!t]
\centering
\includegraphics[width=3.5in]{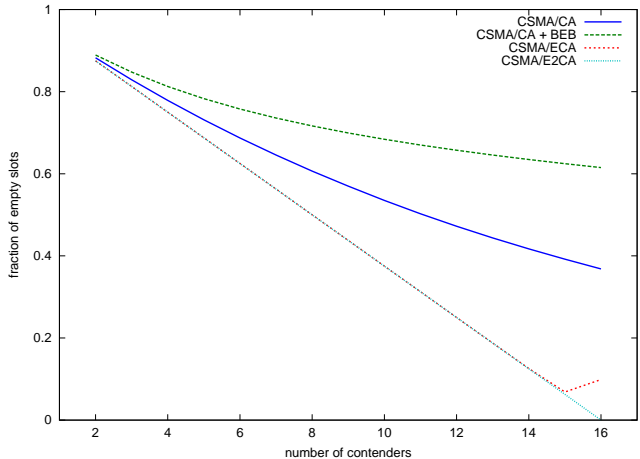}
\caption{Fraction of empty slots for each of the three protocols under comparison for different number of contenders.}
\label{fig:baseline_empty}
\end{figure}

\begin{figure}[!t]
\centering
\includegraphics[width=3.5in]{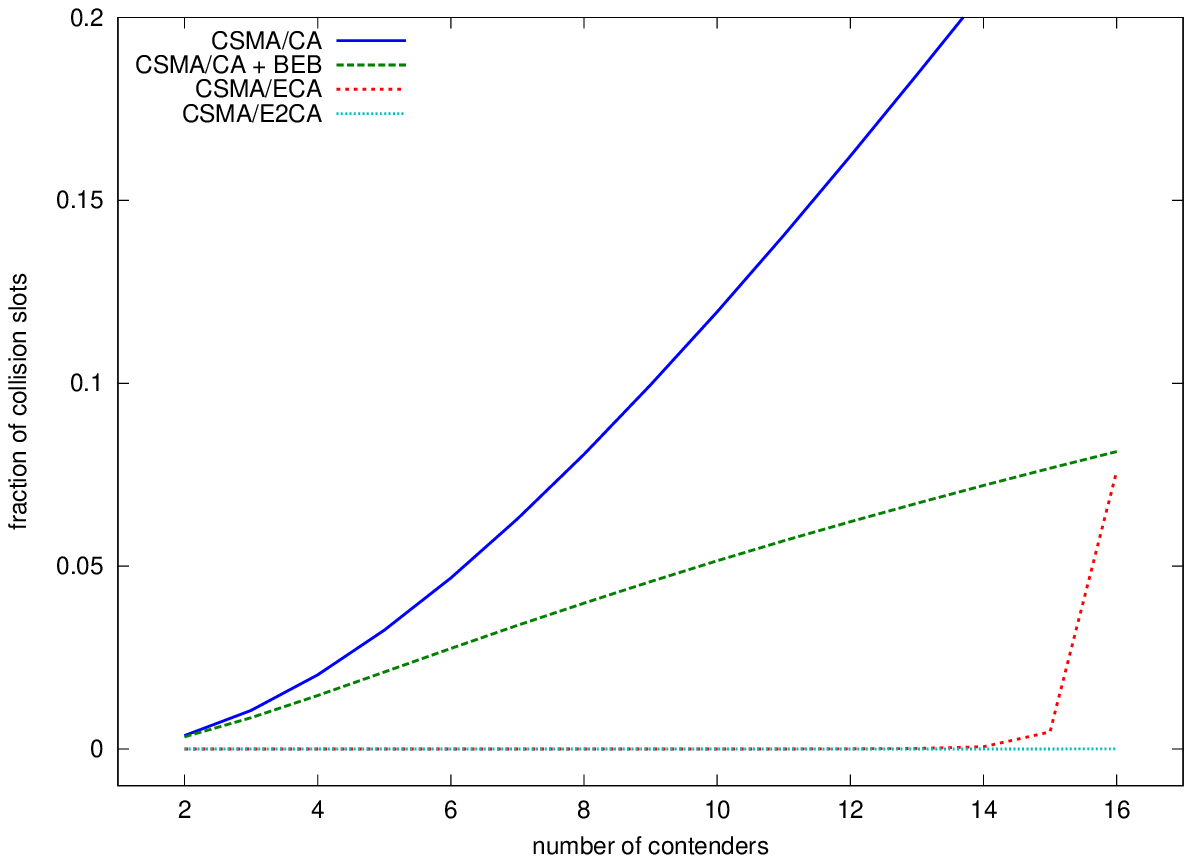}
\caption{Fraction of collision slots for each of the three protocols under comparison for different number of contenders.}
\label{fig:baseline_collision}
\end{figure}

\begin{figure}[!t]
\centering
\includegraphics[width=3.5in]{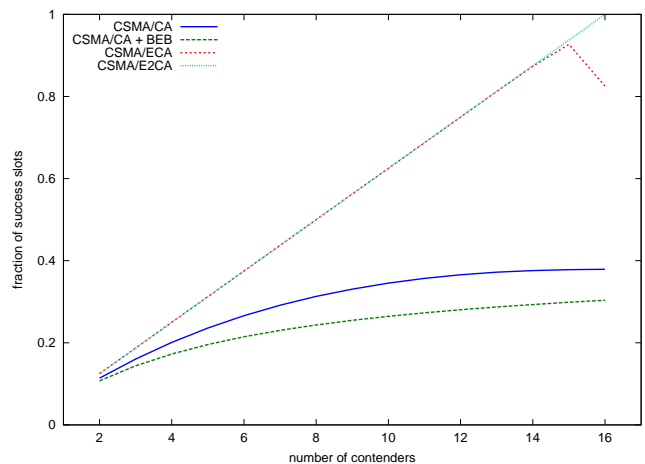}
\caption{Fraction of success slots for each of the three protocols under comparison for different number of contenders.}
\label{fig:baseline_success}
\end{figure}

Fig. \ref{fig:baseline_empty} shows the fraction of empty slots. Empty slots are usually very short and therefore the performance penalty incurred for having large fraction of empty slots is substantially less than the one caused by collisions. Collisions are depicted in Fig. \ref{fig:baseline_collision}. This figure shows that both CSMA/ECA and CSMA/E2CA effectively prevent collisions for a number of contenders lower than $C=16$. For the particular case of a number of contenders equal to $C$, the transient state is very long for CSMA/ECA and therefore we can observe that the number of collisions is relatively high. Finally, the number of successful slots is presented in Fig. \ref{fig:baseline_success}.

From the results it can be observed that CSMA/ECA and CSMA/E2CA consistently outperform the other two protocols. The difference between CSMA/ECA and CSMA/E2CA in ideal channel conditions is obvious only for a large number of contenders.

For the next simulations round we will consider a channel that introduces 10\% packet drop. When a single station transmits in a given slot, that packet is dropped with probability equal to 10\%. The station is not able to differentiate between packet drop and collision, and therefore it will react as if a collision has occurred.

\begin{figure}[!t]
\centering
\includegraphics[width=3.5in]{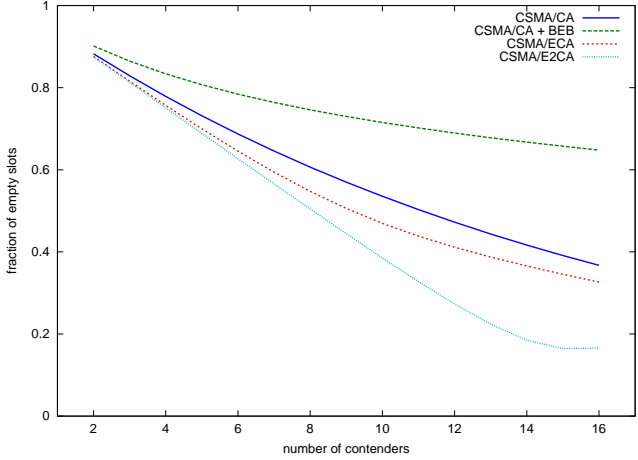}
\caption{Fraction of empty slots for each of the three protocols under comparison for different number of contenders in a channel that drops 10\% of packets.}
\label{fig:with_errors_empty}
\end{figure}

\begin{figure}[!t]
\centering
\includegraphics[width=3.5in]{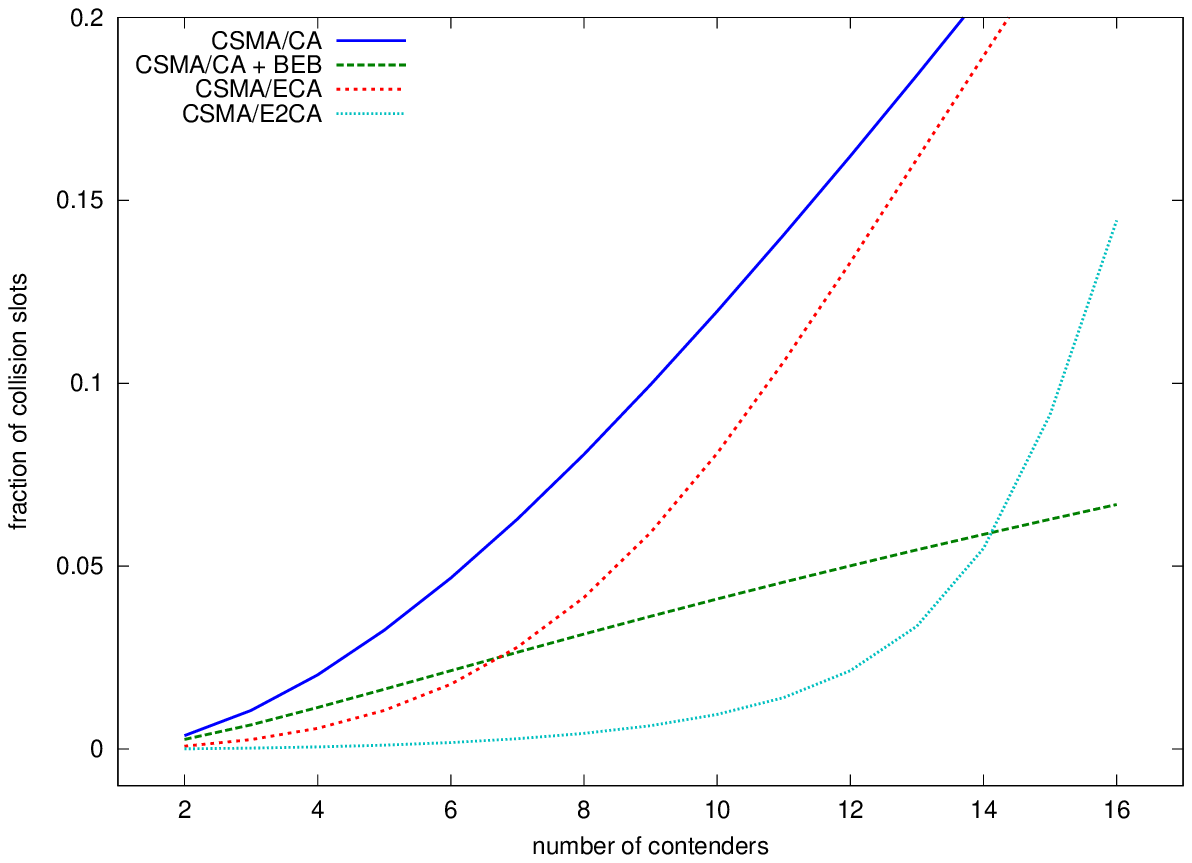}
\caption{Fraction of collision slots for each of the three protocols under comparison for different number of contenders in a channel that drops 10\% of packets.}
\label{fig:with_errors_collision}
\end{figure}

\begin{figure}[!t]
\centering
\includegraphics[width=3.5in]{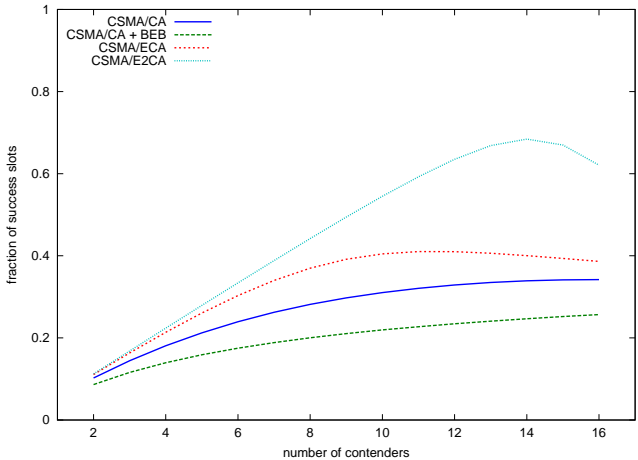}
\caption{Fraction of success slots for each of the three protocols under comparison for different number of contenders in a channel that drops 10\% of packets.}
\label{fig:with_errors_success}
\end{figure}

Fig. \ref{fig:with_errors_empty} shows the fraction of empty slots of the four protocols in a lossy channel. Fig. \ref{fig:with_errors_collision} shows the fraction of collisions and Fig. \ref{fig:with_errors_success} the fraction of successful transmissions. The fraction of packet drops is not shown, since it is equal to a 10\% of successful transmissions. 

From the results we can conclude that CSMA/CA presents a behaviour that is very similar to the one we have observed in the ideal channel. This is not the case of CSMA/ECA that presents an impaired behaviour because it cannot differentiate between channel errors and collisions. This problem is partly alleviated by CSMA/E2CA. Nevertheless, if we want to obtain collision-free operation after a short transient state, the number of competing stations has to be around one half of the capacity of the system. As an example, for a deterministic value after successes $C=16$, we can accommodate up to $\sigma=8$ stations before the number of collisions starts to grow. If we expect that a number of stations much larger than 8 is going to be simultaneously backlogged (with a packet ready to send), then mechanisms that dynamically adjust the contention parameter should be considered. This is the subject ot the next section.

\section{Dynamic Parameter Adjustment in CSMA/E2CA}
\label{sec:dpa}
In the previous sections we have studied the performance of CSMA/ECA in scenarios in which the number of active contenders was below the deterministic backoff value used after successful transmissions. This is the situation that is most often encountered in current deployments, in which the number of stations registered to a single access point is moderate and only a fraction of those stations are contending for the channel.

Nevertheless, it would be desirable that a contention protocol for future WLANs was scalable enough to accommodate an arbitrarily large number of contenders. In the present section we discuss the dynamic parameter adjustment of CSMA/CA to accommodate a changing (and potentially high) number of contenders.

The idea of adjusting the $CW_{min}$ parameter in CSMA/ECA was initially presented in \cite{barcelo2010dpa}. In the current paper we review that initial approach, justify its convenience, and present additional results. Our approach relies on the existence of a central entity (access point) that can detect whether the channel slots are empty or busy and computes and distributes the value for $CW_{min}$. There are alternative proposals that do not require the existence of such central entity and therefore are appropriate for distributed scenarios \cite{fang2009dlm}.

In an infrastructure deployment, the access point sends beacon intervals approximately every 100 ms. This beacon frames include control information such as the $CW_{min}$ to be used by the contending stations. For convenience, we will define a \emph{beacon interval} as the time that elapses between two beacon transmissions. In our approach, we adjust the $CW_{min}$ in such a way that the fraction of busy slots ($\beta$) between two beacon frames is a value around $\beta_{target}= \frac{1}{4}$. 

In fact, we accept that the fraction of busy slots takes values between $\frac{1}{8}$ and $\frac{1}{2}$. These particular values are tied to the fact that, according to the standard, the minimum contention window has to be a power of two (minus one). If the fraction of busy slots exceeds $\frac{1}{2}$ (i.e., it takes a value between $\frac{1}{2}$ and $1$), then the doubling of the contention window would desirably halve the fraction of the busy slots to a value between $\frac{1}{4}$ and $\frac{1}{2}$, which is within our acceptable range.

 The wide range of acceptable values provides an hysteresis margin that prevents the oscillation of the $CW_{min}$ value in use. Table 2 in \cite{barcelo2010dpa} provides an example that shows that even when collision-free operation is reached, the exact number of success, collision and empty slots may change from one beacon interval to the next. The hysteresis margin prevents that this little changes triggers an update of the $CW_{min}$ value. Notice that decreasing the $CW_{min}$ value may originate several collisions and move the system from the steady-state collision-free operation to the transient state. For this reason the reduction of the $CW_{min}$ value should be prevented unless it is strictly necessary (i.e., unless the fraction of busy slots is very low). 

The recursive equations that regulate the behaviour of the system as described in the previous paragraphs are the following:
\begin{equation}
\begin{array}{lr}
CW_{min,i}=CW_{min}^{default} ,&  \; i=0 \\
CW_{min,i}=\\
\max (CW_{min}^{default}, CW_{min,i-1} \cdot 2^{trunc (\log_2 \frac{\beta}{\beta_{target} } )} ) & \; i>0
\end{array}
\label{eq:update}
\end{equation}
where $i$ is an index on the beacon interval and $CW_{min, i}$ is the minimum contention window used in the $i$-th beacon interval.

The minimum contention window is initially set to its default value. Then, after 100 ms, the  new minimum contention window is computed as a function of the previous value and the measured fraction of busy slots during the last beacon interval. The new value for $CW_{min}$ is conveyed in the subsequent beacon frame. As in \cite{patras2009cta}, we take the design decision to prevent that the minimum contention window takes values below the default value defined in the standard.

In (\ref{eq:update}) we make use of the truncate ($trunc(\cdot)$) function that takes the integer part of its argument and discards the decimals. For implementation simplicity, it would be desirable to avoid floating-point operations. The following pseudo-code can be used to adjust the value of $CW_{min}$ using integer arithmetic. We double $CW_{min}$ when the fraction of busy slots is large and we halve $CW_{min}$ if the measured fraction is small.

\begin{algorithm}
  \tcc{Initialize $CW_{min}$.}
  $CW_{min} \leftarrow CW_{min}^{default}$\;
  \While{true}{
    \tcc{Measure the number of empty and busy slots in the beacon interval}
    measure($empty$,$busy$)\;
    \If{$8*busy < (empty+busy)$}{
      \If{$CW_{min}>CW_{min}^{default}$}{
        $CW_{min} \leftarrow CW_{min} / 2$\;
      }
    }
    \If{$busy > empty$}{
      $CW_{min} \leftarrow CW_{min} \cdot 2$\;
    }
    
  }
  \caption{$CW_{min}$ adaption in CSMA/ECA.}
\label{alg:adaption}
\end{algorithm}

Our suggested mechanism does not aim at filling every slot with a successful transmission. On the contrary, we advocate for leaving a large fraction of slots empty. The fraction of empty slots in steady state operation will be a number between $\frac{1}{2}$ and $\frac{7}{8}$. In the results presented at the end of this section we will see that this large number of empty slots does not severely penalize performance. The rationale for leaving a large number of slots empty is twofold. First, it guarantees quick converge to collision-free operation after any disrupting event, such as an interference burst. Second, the fraction of empty slots would prevent the starvation of legacy terminals if they were present. 

As in \cite{barcelo2010dpa}, we will stress our protocol by simulating a scenario in which a large number of contenders simultaneously join the network with ideal channel conditions. In \cite{barcelo2010dpa} we presented results for a number of contenders equal to 20. In the present work we present results for various numbers of contenders up to 100. There are two additional differences to take into account when comparing the present results with our previous work. In the present paper we have used CSMA/E2CA and we have set $CW_{max}=CW_{min}$, while in \cite{barcelo2010dpa} we used CSMA/ECA and $CW_{max} > CW_{min}$.

\begin{figure}[!t]
\centering
\includegraphics[width=3.5in]{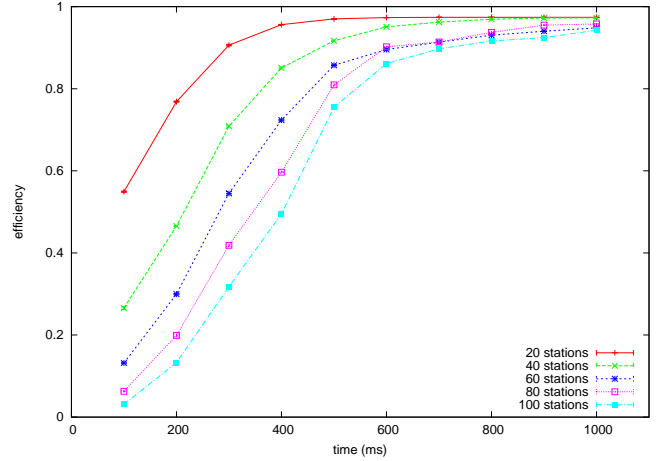}
\caption{Efficiency simulation results when a large number of contenders simultaneously join the network.}
\label{fig:efficiency_one_second}
\end{figure}

The results are presented in Fig. \ref{fig:efficiency_one_second}. We plot the channel efficiency in each of the first ten beacon intervals of operation. Channel efficiency is defined as the fraction of time devoted to successful packet transmissions. Since we are evaluating the contention mechanism, the only inefficiencies we take into account are empty and collision slots. 

For these experiments we have assumed that the stations are saturated. They transmit data to the access point using IEEE 802.11b 11Mbps data rate. The size of data packets is 1500 bytes and all the contenders simultaneously join the network at time 0. We take measures every beacon interval (100ms) and plot average values across 5000 simulations. The 95\% confidence intervals are plotted in the figure, even though they are so small that they are hardly visible.

It can be observed in the figure that our suggested approach quickly reacts and adapts the contention parameter to deliver high performance in less than a second, regardless of the number of simultaneous contenders. A careful observer will detect some crossings between the lines in the figure. The explanation lies in the hysteresis margins that we use: The fraction of empty slots are not necessary equal for the different number of contenders.

It is worth noting that the channel efficiency of CSMA/E2CA with parameter adjustment can be increased to any value arbitrarily close to one. This can be achieved by increasing the length of successful slots. Since the length of empty slots is constant and there are no collision slots after the transient state, the fraction of time devoted to successful transmissions will increase as we increase the duration of a successful transmission. This can be readily achieved using the frame aggregation tools provided by the IEEE 802.11n standard amendment \cite{IEEE80211n-IEEESTD2009}.

\section{Degree of stickiness}
\label{sec:stickiness}
Throughout the paper we have considered the CSMA/E2CA protocol that uses a deterministic backoff for two consecutive times after each successful transmission. We can define the degree of stickiness as the number of deterministic backoffs used after each successful transmission. As an example, the degree of stickiness of CSMA/E2CA is two. It is reasonable to consider the possibility of using a deterministic backoff for several (even infinite) times after each successful transmission. Let us discuss the drawbacks of using a large degree of stickiness. 

Even in the idealistic scenario considered in the current paper, two stations that have successfully transmitted in their last attempt can collide in their following attempt if there has been a reduction of the $CW_{min}$ value. The $CW_{min}$ value may be reduced if there is a large number of stations that leave the contention, as it has been explained in the previous section. If such reduction occurs, two stations using a large degree of stickiness may suffer a large number of collisions before switching to a random behaviour to search for a collision-free schedule.

In a more realistic scenario, clock drifts may occur. As a result, the different stations may no longer be perfectly synchronized and may not decrement their counters exactly at the same time. If this is the case, two stations that have successfully transmitted in their last transmission attempt can still suffer a collision in their next transmission attempt. Again, if the network uses a large degree of stickiness, it will take several collisions for the stations to switch to a random behaviour to look for a collision-free schedule.

\section{Conclusion}
\label{sec:conclusion}

In this paper we have reviewed previous work on learning MAC protocols and, in particular, on CSMA/ECA. We have presented a model that computes the number of slots required to reach collision-free operation, and we have introduced a modification that results in a shorter transitory. In particular, we advocate for the use of a deterministic backoff for two consecutive times after each successful transmission and we name this improved protocol CSMA/E2CA. Then we present extensive simulation results that highlight the benefit of using CSMA/E2CA, specially in those channels that are prone to packet errors.

With static parameter configuration, there is a limit on the number of contenders that can operate in a collision-free fashion in a CSMA/E2CA network. We revisit our previous work on dynamic parameter adjustment for CSMA/ECA to offer a centralized and easy to implement parameter adjustment mechanism for CSMA/E2CA. Finally, we run simulations in which a large number of contenders simultaneously join the network and show that our approach quickly reacts to adjust the contention parameter and deliver high performance for any number of simultaneous contenders.
\section*{Acknowledgment}
The present paper is the extended version of a MACOM workshop paper. The authors benefited from the feedback and discussion in the workshop. The authors are specially indebted with N. Abramson for his clarifications regarding Aloha and with D. Malone for providing recent and valuable references.

This work has been partially supported by the Spanish Government
under projects TEC2008-0655/TEC (GEPETO, Plan Nacional I+D) and CSD2008-00010 (COMONSENS, Consolider-Ingenio Program).


\ifCLASSOPTIONcaptionsoff
  \newpage
\fi



\bibliographystyle{IEEEtran}
\bibliography{IEEEabrv,barcelo2011tcf}
\end{document}